\documentclass[11pt,a4paper]{article}
\usepackage[utf8]{inputenc}
\usepackage{amsmath,amssymb,graphicx,amsthm,dsfont}
\usepackage{color,soul}
\usepackage{xspace}
\usepackage{enumerate}
\usepackage{bbm}
\usepackage{mathtools}
\usepackage{comment}
\usepackage{paralist}
\usepackage{nicefrac}
\usepackage{booktabs}
\usepackage{multirow}

\usepackage{subcaption}
\usepackage[noend,ruled,linesnumbered]{algorithm2e}
\usepackage{xifthen}

\usepackage{geometry}

\usepackage{fullpage}

\usepackage{colortbl}

\usepackage[%
pagebackref,
breaklinks=true,
colorlinks=true,
linkcolor=blue,
citecolor=darkgreen,
urlcolor=purple,
pdfpagelayout=SinglePage,
pdfstartview=Fit]{hyperref}
\renewcommand*{\backref}[1]{}
\renewcommand*{\backrefalt}[4]{%
\ifcase #1%
\marginpar{\tiny no cite}
\or
 $\rightarrow$~p.~#2.%
\else
  $\rightarrow$~pp.~#2.%
\fi
}
\usepackage[numbers]{natbib}

\usepackage{tikz}
\usetikzlibrary{decorations,arrows,petri,topaths,backgrounds,shapes,positioning,fit,calc,decorations.pathreplacing,patterns,intersections,decorations.pathmorphing,matrix}

\makeatletter
\newcommand{\gettikzxy}[3]{%
  \tikz@scan@one@point\pgfutil@firstofone#1\relax
  \edef#2{\the\pgf@x}%
  \edef#3{\the\pgf@y}%
}
\makeatother

\usepackage{cleveref}

\theoremstyle{definition}

\crefname{table}{Table}{Tables}
\crefname{figure}{Figure}{Figures}
\crefname{theorem}{Theorem}{Theorems}
\crefname{definition}{Definition}{Definitions}
\crefname{corollary}{Corollary}{Corollaries}
\crefname{observation}{Observation}{Observations}
\crefname{lemma}{Lemma}{Lemmas}
\crefname{example}{Example}{Examples}
\crefname{reduction}{Reduction}{Reductions}
\crefname{construction}{Construction}{Constructions}
\crefname{subsection}{Subsection}{Subsections}
\crefname{section}{Section}{Sections}
\crefname{proposition}{Proposition}{Propositions}
\crefname{algorithm}{Algorithm}{Algorithms}
\crefname{drule}{Rule}{Rules}
\crefname{claim}{Claim}{Claims}

\newcommand{\SM}{\textsc{Stable Marriage}\xspace}
\newcommand{\SMI}{\textsc{Stable Marriage with Incomplete Preferences}\xspace}
\newcommand{\SMTI}{\textsc{Stable Marriage with Ties and Incomplete Preferences}\xspace}
\newcommand{\SMT}{\textsc{Stable Marriage with Ties and Complete Preferences}\xspace}
\newcommand{\SR}{\textsc{Stable Roommates}\xspace}
\newcommand{\SRI}{\textsc{Stable Roommates with Incomplete Preferences}\xspace}
\newcommand{\SRTI}{\textsc{Stable Roommates with Ties and Incomplete Preferences}\xspace}
\newcommand{\SRT}{\textsc{Stable Roommates with Ties and Complete Preferences}\xspace}
\newcommand{\SRB}{\textsc{Min-Block-Pairs Stable Roommates}\xspace}
\newcommand{\SRD}{\textsc{Min-Delete Stable Roommates}\xspace}
\newcommand{\SRA}{\textsc{Min-Block-Agents Stable Roommates}\xspace}
\newcommand{\SRM}{\textsc{Max-Card Stable Roommates}\xspace}
\newcommand{\SRMin}{\textsc{Min-Card Stable Roommates}\xspace}
\newcommand{\SMM}{\textsc{Max-Card Stable Marriage}\xspace}

\newcommand{\ESR}{\textsc{Egalitarian Stable Roommates}\xspace}
\newcommand{\RSR}{\textsc{Min Regret Stable Roommates}\xspace}

\newcommand{\SESM}{\textsc{Sex-Equal Stable Marriage}\xspace}
\newcommand{\BSM}{\textsc{Balanced Stable Marriage}\xspace}
\newcommand{\ESM}{\textsc{Egalitarian Stable Marriage}\xspace}
\newcommand{\RSM}{\textsc{Min Regret Stable Marriage}\xspace}
\newcommand{\MESM}{\textsc{Min Man-Exchange Stable Marriage}\xspace}

 \newcommand{\myemph}[1]{{\color{winered}\emph{#1}}}

\newcommand{\bp}{\ensuremath{\beta}} % #blocking pairs
\newcommand{\ba}{\ensuremath{\eta}} % #blocking agents
\newcommand{\sss}{\ensuremath{\tau}} % #size of the stable matching
\newcommand{\unmatched}{\ensuremath{\mathsf{n}_{\mathsf{u}}}} % #size of the stable matching
\newcommand{\da}{\ensuremath{\delta}} % #deleted agents
\newcommand{\egalcost}{\ensuremath{\gamma}}
\newcommand{\regretcost}{\ensuremath{\lambda}}
\newcommand{\secost}{\ensuremath{\rho}}
\newcommand{\balcost}{\ensuremath{\alpha}}
\newcommand{\excost}{\ensuremath{\theta}}
\newcommand{\egalcostn}{egalitarian cost}
\newcommand{\Egalcostn}{Egalitarian cost}
\newcommand{\regretcostn}{regret cost}

\newcommand{\po}{\ensuremath{\phi}}

\usepackage{xcolor}
\definecolor{dargray}{rgb}{0.18, 0.18, 0.18}
\definecolor{darkgreen}{rgb}{0.01,0.6,0.1}
\definecolor{lightrose}{rgb}{0.996,0.75,0.793}
\definecolor{rose}{cmyk}{0.75, 0.75, 0,0}
\definecolor{winered}{rgb}{0.6,0.1,0.1}
\definecolor{darkyellow}{rgb}{.99, .87, 0.04}
\definecolor{lightyellow}{rgb}{1, 1, 0.6}
\definecolor{transparent}{rgb}{1,1,1}
\definecolor{lightlightgray}{rgb}{0.88, 0.88, 0.88}
\definecolor{lightgray}{rgb}{0.8, 0.8, 0.8}
\definecolor{lightblue}{rgb}{0.527,0.805,0.977}
\definecolor{lightgreen}{rgb}{.74,1,0}
\usepackage{graphicx}

\DeclareMathOperator{\rank}{rank}

\newcommand{\ecost}{\textsf{egal-cost}}
\newcommand{\rcost}{\textsf{regret-cost}}

\tikzstyle{blueline} = [thick, blue, dotted]
\tikzstyle{redline} = [thick, red, dashed]
\tikzstyle{blackline} = [thick, black]

\title{Computational Complexity of Stable Marriage and Stable Roommates and Their Variants}

\author{\textsc{Jiehua Chen}\\
  University of Warsaw}

\begin{document}

\maketitle

This paper gives an overview on and summarizes existing complexity and algorithmic results of some variants of the \SM and the \SR problems.
The last section defines a list of stable matching problems mentioned in the paper.  
If you find any corrections, suggestions, new or missing results, please send them to \url{jiehua.chen2@gmail.com}.

  \medskip

\section{\SM and \SR}

\SR deals with the division of human or non-human agents (classically called roommates)
into pairs,
where each agent has a preference list on who is more preferable as a partner.
In \SR, we are given a set of agents~$U$,
each having a preference list~$\succ$ (a linear order on $U$),
indicating which other it prefers to have as partner,
The goal of \SR is to search for a \myemph{matching}, \emph{i.e.}\ a set of disjoint pairs of $U$,
that contains \emph{no} blocking pairs. 
Herein, a \myemph{blocking pair} is a pair of two agents $u$ and $w$ who are not matched together such that
\begin{compactenum}[(i)]
  \item  $u$ is either unmatched by $M$ or prefers $w$ to its partner~$M(u)$, \emph{i.e.}\ $w\succ_u M(u)$, and
  \item $w$ is either unmatched by $M$ or prefers $u$ to its partner~$M(w)$, \emph{i.e.}\ $u \succ_w M(w)$.
  \end{compactenum}
We call such a matching a \myemph{stable matching} and we call it \myemph{perfect} if every agent is assigned a partner.
An even more well-known problem---\SM, is a bipartite, restricted variant of \SR, where the agents are partitioned into two disjoint subsets such that each agent has preferences over the agents of the opposite subset and can only be assigned a partner from that subset.
\SM and \SR are introduced in the field of Economics and Computer Science in the 1960s and 1970s, respectively~\cite{GaleShapley1962,Knuth1976}.
They have since been studied extensively by not only economists but also computer scientists and social and political scientists~\cite{Knuth1976,Irving2016a,Irving2016b,Manlove2008,GusfieldIrving1989,Manlove2013,AbBiMa2005}.

Practical applications of \SR as well as \SM (and their variants)
include partnership issues in the real-world~\cite{GaleShapley1962},
allocating items (e.g.\ time-slots, locations, or resources) to agents (e.g.\ event hosts or individuals), where the agents have preferences over the available items~\cite{HyZe1979,AbrCheKummir2006,ChenSoenmez2002},
centralized automated mechanisms that assign children to schools ~\cite{AbPaRo2005a,AbPaRo2005b}, school graduates to universities~\cite{BaBa2004,BiKi2015}, or medical students to hospitals~\cite{NResidentMatchingP,SResidentMatchingP}, grouping people to share a room or pairing participants for chess tournaments or in P2P networks ~\cite{KuLiMa1999,LeMaViGaReMo2006,GaLeMamoReVi2007},
scheduling user jobs to machines so that users do not want to switch to some other machines, and finding receiver-donor pairs for organ transplants (for more details, see the work of \citet{MaOM2014} and \citet{RoSoUn2007,RoSoUn2005}.

\subsection{When preferences are beyond complete linear orders}
Varying from application to application, the preferences of the agents in \SR and \SM could be \myemph{incomplete}, meaning that not every agent is an acceptable partner to every agent of the opposite sex. 
This corresponds to the case where the underlying \myemph{acceptance graph} is bipartite yet incomplete.
The preferences of the agents may also contain \myemph{ties},
meaning that two agents are considered to be equally good as a partner.
Note that for preferences with ties, our stability concept is regarded as \emph{weak} stability since it suffices to forbid the existence of unmatched pairs that \myemph{strictly} prefer to be with each other.
There are two other popular stability concepts--strong stability and super stability.
A matching is called \myemph{strongly stable} if no unmatched pair~$\{u,w\}$ exists such that
\begin{inparaenum}[(i)]
  \item $u$ is either unmatched or finds $w$ at least as good as its assigned partner~$M(u)$ (denoted as $w \succeq_u M(u)$), and
  \item $w$ is either unmatched or prefers~$u$ to its assigned partner~$M(w)$ (denoted as $u \succ_w M(w)$).
\end{inparaenum}
A matching is called \myemph{super stable} if no unmatched pair~$\{u,w\}$ exists such that
\begin{inparaenum}[(i)]
  \item $u$ is either unmatched or finds $w$ at least as good as its assigned partner~$M(u)$ (denoted as $w \succeq_u M(u)$), and
  \item $w$ is either unmatched or finds $w$ at least as good as its assigned partner~$M(w)$ (denoted as $u \succeq_w M(w)$).
\end{inparaenum}
We refer to the works of \citet{GusfieldIrving1989,Manlove2013} for more detailed discussions.
In this paper, unless stated explicitly, we always write stable matchings to refer to weakly stable matchings. 

For preferences without ties, even when they are incomplete,
deciding whether a \SRI instance with $n$ agents admits a stable matching
and finding one if it exists can be done in $O(n^2)$ time, for both \SR and \SM~\cite{GaleShapley1962,Irving1985}.
However, when preferences have ties, \SR becomes NP-hard~\cite{Ronn1990} even if the preferences are complete.
The situation for \SM seems more positive: It still admits a stable matching, even for incomplete preferences.
But, there could be stable matchings with different sizes.
By breaking the ties arbitrarily, one can use the algorithm of \cite{Irving1994} to find a stable matching.

The corresponding computational problems are defined as {\SRI~\normalfont{(SRI)}} and {\SMI~\normalfont{(SMI)}}, {\SRT~\normalfont{(SRT)}} and {\SMT~\normalfont{(SMT)}}, and {\SRTI~\normalfont{(SRTI)}} and {\SMTI~\normalfont{(SMTI)}}.

\subsection{When preferences satisfy structural properties}

The original stable roommate problem strives to pair up roommates who prefer to stay with each other. 
In this setting, the roommates' preferences over who is more suitable to stay with may display some specific structure.
For instance, from the psychological point of view, it is natural to assume that individuals are \myemph{narcissistic}, meaning that 
they like to be with someone who resembles themselves.
Furthermore, as different individuals may e.g. have different ideal room temperatures, the preferences of an individual may display the so-called \emph{single-peaked property}: individuals will prefer those whose ideal room temperature is close to their own over those whose temperature of choice is further away.

\citet{BartholdiTrick1986} showed that when the input preferences are linear, narcissistic (i.e.\ each agent ranks itself at the first position),
and single-peaked
(i.e.\ there is a linear order of the agents such that each agent's preferences over the agents along this order are either strictly increasing, strictly decreasing, or first strictly increasing and then strictly decreasing), 
then a \SR (SR) instance with $2n$ agents always admits a unique stable matching and it can be found in sublinear time~($O(n)$).
\citet{BreCheFinNie2017} showed that the same holds when the given instance is linear, complete, narcissistic, and single-crossing (i.e.\
there is a linear order of the agents such that for each pair of agents~$\{x,y\}$, there are at most two consecutive agents along the order that disagree over the relative order of $x$ and $y$).

When ties are present but the preferences remain complete, \citet{BreCheFinNie2017} managed to show that the in general NP-hard \SRT problem~\cite{Ronn1990} becomes polynomial-time solvable.
However, when incompleteness also comes into play, \SRT remains NP-hard even if the input preferences are narcissistic, single-peaked, and single-crossing.

\cref{tab:results-sm,tab:results-sr} summarize results of (variants of) \SM{} and \SR{} for two aspects, without and with ties.

Besides the relevant work mentioned in the paper, 
there are some other research regarding the parameterized complexity of preference-based stable matching problems~\cite{GupSauZeh2017,CheNieSko2018,CheSkoSor2019}.
 
\newcommand{\mynew}[1]{{\color{winered}\textbf{#1}}}

\newcommand{\citeronn}{$^\triangle$\xspace}
\newcommand{\citeirv}{$^\diamondsuit$\xspace}
\newcommand{\citegi}{$^\spadesuit$\xspace}
\newcommand{\citefeder}{$^\heartsuit$\xspace}
\newcommand{\citechsy}{$^\star$\xspace}
\newcommand{\other}[1]{{\color{gray}#1}}
\newcommand{\citeabm}{$^{\bullet}$\xspace}
\newcommand{\citeirvtwo}{$^{\clubsuit}$\xspace}
\newcommand{\citemiimm}{$^{\square}$\xspace}
\newcommand{\citecim}{$^{+}$\xspace}
\newcommand{\citetan}{$^{\Join}$\xspace}

%%%%%%%%%%%%%%%%%%%%%%%%%%%%%%%%%%%%%%%%%%%%%%%%%%%%%%
%%%%%%%%%%%%%%%%%%%%%%% SM %%%%%%%%%%%%%%%%%%%%%%%%%%%%%
%%%%%%%%%%%%%%%%%%%%%%%%%%%%%%%%%%%%%%%%%%%%%%%%%%%%%%
\renewcommand{\cellcolor}[1]{}
\begin{table}
\centering
\resizebox{\textwidth}{!}{
  \begin{tabular}{@{}l@{\qquad}l@{\;}c@{}c@{}c@{}}
    \toprule
    \multicolumn{5}{c}{Marriage Setting}\\
    \midrule
    \multicolumn{2}{l}{Objective} & Without ties &~~~& With ties\\
    & Parameter &    & & \\
    \midrule
    \multicolumn{2}{l}{Any Stable} & \multicolumn{1}{c}{$O(n^2)$~\cite{Irving1985,GusfieldIrving1989}} && \multicolumn{1}{c}{$O(n^2)$~\cite{Irving1994}}   \\
    \\[-1ex]
    
   \multicolumn{2}{l}{\textsc{Max Cardinality}} %\\
              & \multicolumn{1}{c}{-} && \multicolumn{1}{c}{NP-c (perfect)~\cite{MaIrIwMiMo2002}}\\
      & \cellcolor{lightgray}Max.\ list length~$\ell$ & \multicolumn{1}{c}{\cellcolor{lightgray}-} &\cellcolor{lightgray}& \multicolumn{1}{c}{\cellcolor{lightgray}NP-c~($\ell=3$)~\cite{IrvManOMa2009}}\\
    & Cardinality~\sss &  \multicolumn{1}{c}{-} &&\multicolumn{1}{c}{$O(\sss^{\sss})$, size-$O(\sss^2)$~kern~\cite{AdGuRoSaZe2018}}\\
    & \cellcolor{lightgray}Unmatched agents~$\unmatched=2n - 2\sss$ &  \multicolumn{1}{c}{\cellcolor{lightgray}-}&\cellcolor{lightgray}& \multicolumn{1}{c}{\cellcolor{lightgray}NP-c ($\unmatched=0$)~\cite{MaIrIwMiMo2002}}\\
    \\[-1ex]
     
      \multicolumn{2}{l}{\textsc{Min Cardinality}} %\\
                & \multicolumn{1}{c}{-} && \multicolumn{1}{c}{NP-c~\cite{MaIrIwMiMo2002}}\\
                                  & \cellcolor{lightgray}Matching cardinality~$\hat{\sss}$ &  \multicolumn{1}{c}{\cellcolor{lightgray}-} &\cellcolor{lightgray}&\multicolumn{1}{c}{\cellcolor{lightgray}$O(\hat{\sss}^{\hat{\sss}})$, size-$O(\hat{\sss}^2)$~kern~\cite{AdGuRoSaZe2018}}\\
    & Unmatched agents~$\unmatched=2n - 2\hat{\sss}$ &  \multicolumn{1}{c}{-}&& \multicolumn{1}{c}{?}\\

    \midrule
    
   \multicolumn{2}{l}{\textsc{Egalitarian}} &
            \multicolumn{1}{c}{P~\cite{IrLeGu1987,Gusfield1987}} &&  \multicolumn{1}{c}{NP-c (even for compl. prefs)~\cite{MaIrIwMiMo2002}}\\
     & Egal.\ cost~$\egalcost$ $=$ pref.\ list length &  \multicolumn{1}{c}{-}& &\multicolumn{1}{c}{$\egalcost^{O(\egalcost)}\cdot n^{O(1)}$~\cite{CHSYicalp-par-stable2018}}\\
    % \multicolumn{1}{c}{\mynew{W[1]-h (?!)}\citechsy}\\
    & Max.\ list length~$\ell$&  \multicolumn{1}{c}{-} & &  \multicolumn{1}{c}{NP-c ($\ell=5$)~\cite{CHHY18wp}} \\
    \\[-1ex]

    \multicolumn{2}{l}{\textsc{Min Regret}} &
                         \multicolumn{1}{c}{P \cite{IrLeGu1987,Gusfield1987}}& &  \multicolumn{1}{c}{NP-c \cite{MaIrIwMiMo2002,HaIrIwMaMiMoSc2003}}\\
    & Max.\ list length $\ell$ &  \multicolumn{1}{c}{-}&& \multicolumn{1}{c}{NP-c ($\ell = 3$) \cite{CsIrMa2016}}\\
    & \regretcostn~$\regretcost$ &  \multicolumn{1}{c}{-}&& \multicolumn{1}{c}{NP-c ($\regretcost = 2$, perfect)~implied by~\cite{IrvManOMa2009}}\\
     & \regretcostn~$\regretcost$           &  \multicolumn{1}{c}{-}&& \multicolumn{1}{c}{? for any stable matching}\\\\[-1ex]
    
    \multicolumn{2}{l}{\textsc{Sex-Equal}} & \multicolumn{1}{c}{NP-c~\cite{Kato1993}} &&  \multicolumn{1}{c}{NP-c~\cite{Kato1993}} \\
    & Sex-equal cost~$\secost$ & \multicolumn{1}{c}{NP-c ($\secost=0$, $\ell=3$)~\cite{MDIrv2014}} &&  \multicolumn{1}{c}{NP-c ($\secost=0$, $\ell=3$)~\cite{MDIrv2014}} \\
    & Max.\ list length~$\ell$ & \multicolumn{1}{c}{$O(n^3)$ ($\ell\le 2$ on one side)~\cite{MDIrv2014}} &&  \multicolumn{1}{c}{?} \\\\[-1ex]

    \multicolumn{2}{l}{\textsc{Balanced}} & \multicolumn{1}{c}{NP-c~\cite{Feder1992a}} &&  \multicolumn{1}{c}{NP-c~\cite{Feder1992a}} \\
         & Balance cost~$\balcost$   & \multicolumn{1}{c}{FPT~(implied by~\cite{GuRoSaZe2017})} && \multicolumn{1}{c}{?}\\
    \\[-1ex]
    \multicolumn{2}{l}{\textsc{Min Man-Exchange}} & \multicolumn{1}{c}{NP-c~\cite{Irving2008}} & &\multicolumn{1}{c}{NP-c~\cite{Irving2008}} \\
    & Exchanges~$\excost$ & \multicolumn{1}{c}{NP-c~($\excost=0$)~\cite{Irving2008}} &&  * \\
    & Max.\ list length~$\ell$ & \multicolumn{1}{c}{P~($\ell=2$)~\cite{Irving2008}, NP-c~($\excost=0$, $\ell=4$)~\cite{DerCheSu2007}} &&  * \\    \\[-1ex]

    % \textsc{Divorce} &
    %             \multicolumn{1}{c}{\mynew{NP-c (wp)}}  & &  \multicolumn{1}{c}{\mynew{NP-c (wp)}}\\

    % \\[-1ex]

    \multicolumn{2}{l}{\textsc{Popular}} &
                       \multicolumn{1}{c}{P~\cite{Gaerdenfors1975}}& &  \multicolumn{1}{c}{NP-c \cite{AbrIrvKavMeh2007}}\\
    & Matching cardinality & \multicolumn{1}{c}{P~\cite{HuangKavitha2013}} &&  \multicolumn{1}{c}{NP-c \cite{AbrIrvKavMeh2007}} \\
   \bottomrule
\end{tabular}
}
\caption{Complexity of the various variants of \SM. ``-'' means that the corresponding parameter is irrelevant for fixed-parameter tractability but may be interesting for the FPT-in-P research.
  ``$*$'' means that the hardness result for the case without ties transfer to the case with ties.} 
%  ``{\color{winered}(wp)}'' means the proof for the corresponding result is provided in some manuscripts not published yet.}
\label{tab:results-sm}
\end{table}

%%%%%%%%%%%%%%%%%%%%%%%%%%%%%%%%%%%%%%%%%%%%%%%%%%%%%%
%%%%%%%%%%%%%%%%%%%%%%% SR %%%%%%%%%%%%%%%%%%%%%%%%%%%%%
%%%%%%%%%%%%%%%%%%%%%%%%%%%%%%%%%%%%%%%%%%%%%%%%%%%%%%

\begin{table}[t!]
\centering
\resizebox{\textwidth}{!}{
  \begin{tabular}{ll@{\;}c@{}c@{}c@{}}
    \toprule
    \multicolumn{5}{c}{Roommates Setting}\\
    \midrule
    \multicolumn{2}{l}{Objective} & Without ties &~~& With ties\\
    \qquad &Parameter &    & & \\
    \midrule
    \multicolumn{2}{l}{Any Stable} & \multicolumn{1}{c}{$O(n^2)$~\cite{Irving1985,GusfieldIrving1989}} && \multicolumn{1}{c}{NP-c~\cite{Ronn1990}} \\
    & max.\ list length~$\ell$ & - && \multicolumn{1}{c}{NP-c~($\ell=3$)~\cite{CsIrMa2016}}\\
    \\[-1ex]
    
    \multicolumn{2}{l}{\textsc{Min Block-Pairs}}%\\
              &  \multicolumn{1}{c}{NP-c \cite{AbBiMa2005}}&&  \multicolumn{1}{c}{NP-c ($\bp=0$)~\cite{Ronn1990}}\\
   & Blocking pairs $\bp$&  \multicolumn{1}{c}{XP~\cite{AbBiMa2005}, W[1]-h ($\ell=5$)~\cite{CHSYicalp-par-stable2018}} & &  \multicolumn{1}{c}{NP-c ($\bp=0, \ell=3$)~\cite{BiMaMcD2012}}\\
 %   \quad $\bp +{}$max.\ list length~$\ell$&  \multicolumn{1}{c}{?} && \multicolumn{1}{c}{NP-c ($\bp=0, \ell=3$)~\cite{BiMaMcD2012}} \\
    \\[-1ex]

  \multicolumn{2}{l}{\textsc{Min Block-Agents}}
    & \multicolumn{1}{c}{NP-c~\cite{CHSYicalp-par-stable2018}} && \multicolumn{1}{c}{NP-c ($\ba=0$)~\cite{Ronn1990}}\\
    &  Blocking agents $\ba$ & \multicolumn{1}{c}{XP, W[1]-h ($\ell=5$)~\cite{CHSYicalp-par-stable2018}} && \multicolumn{1}{c}{NP-c ($\ba=0$)~\cite{Ronn1990}}\\
    \\[-1ex]
    
     \multicolumn{2}{l}{\textsc{Min Delete}}%\\
              &  \multicolumn{1}{c}{P~\cite{Tan1991}}&&  \multicolumn{1}{c}{NP-c ($\da=0$) \cite{Ronn1990}}\\
    & Deleted agents \da &  \multicolumn{1}{c}{-}&&   \multicolumn{1}{c}{NP-c ($\da=0$) \cite{Ronn1990}}\\
    & Remaining agents & \multicolumn{1}{c}{-} & & \multicolumn{1}{c}{FPT~\cite{CHHY18wp}}\\
    \\[-1ex]

    \multicolumn{2}{l}{\textsc{Max Cardinality}} %\\
              & \multicolumn{1}{c}{P \cite{GusfieldIrving1989}} && \multicolumn{1}{c}{NP-c \cite{Ronn1990,MaIrIwMiMo2002}}\\
    & Matching cardinality~\sss &  \multicolumn{1}{c}{-} &&\multicolumn{1}{c}{NP-c ($\sss = 1$) \cite{Ronn1990,MaIrIwMiMo2002}}\\
   & Unmatched agents $\unmatched=n - 2\sss$ &  \multicolumn{1}{c}{-}&& \multicolumn{1}{c}{NP-c ($\unmatched=0$)~\cite{MaIrIwMiMo2002}}\\
    & Size of a max.\ matching &  \multicolumn{1}{c}{-}&&\multicolumn{1}{c}{FPT, poly.\ kern~\cite{AdGuRoSaZe2018}}\\
    \\[-1ex]
    \midrule
    
     \multicolumn{2}{l}{\textsc{Egalitarian}} &
            \multicolumn{1}{c}{NP-c~\cite{Feder1992b}} &&  \multicolumn{1}{c}{NP-c~\cite{Ronn1990}}\\
     & Egalitarian cost~$\egalcost$ &\\             
    & \quad   Unmatched agents' costs & &&\\
    & \;\quad $=$ pref.\ list length&  \multicolumn{1}{c}{$O(2^{\egalcost}\cdot n^2)$, size-$O(\egalcost^2)$ kern.~\cite{CHSYicalp-par-stable2018}}& &\multicolumn{1}{c}{$\egalcost^{O(\egalcost)}\cdot n^{O(1)}$~\cite{CHSYicalp-par-stable2018}}\\
     & \;\quad $=$ a constant &   \multicolumn{1}{c}{$O(2^{\egalcost}\cdot n^2)$, size-$O(\egalcost^2)$ kern.~\cite{CHSYicalp-par-stable2018}}&& \multicolumn{1}{c}{W[1]-h, XP~\cite{CHSYicalp-par-stable2018}}\\
    & \;\quad $= 0$ &   \multicolumn{1}{c}{$O(2^{\egalcost}\cdot n^2)$, size-$O(\egalcost^2)$ kern.~\cite{CHSYicalp-par-stable2018}}& & \multicolumn{1}{c}{NP-c~($\egalcost=0$) ~~\cite{CHSYicalp-par-stable2018}}\\
                                                                                                                                                                                                   % \multicolumn{1}{c}{\mynew{W[1]-h (?!)}\citechsy}\\
    &  Max.\ list length~$\ell$&  \multicolumn{1}{c}{NP-c ($\ell=3$) \cite{CsIrMa2016}} & &  \multicolumn{1}{c}{*} \\
    \\[-1ex]

     \multicolumn{2}{l}{\textsc{Min Regret}} &
                         \multicolumn{1}{c}{P \cite{GusfieldIrving1989}}& &  \multicolumn{1}{c}{NP-c \cite{MaIrIwMiMo2002}}\\
    & Max.\ list length $\ell$ &  \multicolumn{1}{c}{-}&& \multicolumn{1}{c}{NP-c ($\ell = 3$) \cite{CsIrMa2016}}\\
    & \regretcostn~\regretcost &  \multicolumn{1}{c}{-}&& \multicolumn{1}{c}{NP-c ($\regretcost = 2$) implied by \cite{CsIrMa2016}}\\
    \\[-1ex]

    % \textsc{Divorce} &
    %                   \multicolumn{1}{c}{\mynew{NP-c (wp)}}& &  \multicolumn{1}{c}{\mynew{NP-c (wp)}}\\
    % \\[-1ex]

     \multicolumn{2}{l}{\textsc{Popular}} &
                       \multicolumn{1}{c}{NP-c~\cite{GuMiSaZe2019soda,FaeKavPowZha2019soda}}& &  \multicolumn{1}{c}{*}\\
    \bottomrule
\end{tabular}
}
\caption{Complexity of the various variants of \SR. ``-'' means that the corresponding parameter is irrelevant for fixed-parameter tractability but may be interesting for the FPT-in-P research.
  ``$*$'' means that the hardness result for the case without ties transfer to the case with ties.}
%  ``{\color{winered}(wp)}'' means the proof for the corresponding result is provided in some manuscripts not published yet.}
\label{tab:results-sr}
\end{table}

\section{Optimization variants}

As noted in the introduction, some \SR instances do not admit any stable matching at all, and in fact, empirical study suggests that a constant fraction of all sufficiently large instances will have no solution~\cite{PitIrv1994}. Moreover, even if a given \SR instance admits a stable matching, this solution may not be unique, and there might be solutions with which the agents are more satisfied than with others and thus, are more desirable than others.
Given these two facts, it makes sense to consider two types of optimization variants for \SR:
In one type one would want to compute stable matching that optimize a certain social criteria; in the other, one would want to compute matchings with optimal distance or closeness to stability.

\subsection{Socially optimal stable matchings}
For the case when more than one stable matching exists, it is desirable to compute a stable matching that is socially most satisfactory. Herein, the satisfaction of an agent~$x$ with respect to a given stable matching typically depends on the \myemph{rank} of its partner~$y$ assigned by this matching, which is the number of agents that are strictly preferred to $y$ by~$x$.

A stable matching is regarded as socially optimal if
\begin{itemize}
\item[-] the sum of the ranks of all partners (\emph{i.e.}, the \myemph{\egalcostn} $\gamma$) is minimum, or
\item[-] the maximum rank of any partner is (\emph{i.e.}, the \myemph{\regretcostn} $\lambda$) is minimum.
%\item[-] the absolute value of the difference between the sum ranks of one side (\myemph{i.e.}, the \myemph{\partnercostn} $\rho$) is minimum.
\end{itemize}
Accordingly, we define the \myemph{\egalcostn} and the \myemph{\regretcostn} % , and the \emph{\partnercostn} 
of a matching~$M$ as
\begin{align*}
  \ecost(M) \coloneqq& \sum_{\{i,j\}\in M} (\rank_i(j)+\rank_j(i))\text{, and}\\
  \rcost(M) \coloneqq& \max_{i \in V(M)} \rank_i(M(j)).
\end{align*}
%Again, we consider the profiles in \cref{fig:example}. Altough matching~$M_2^3$ has a larger cardinality than $M_2^1$, matching~$M_2^1$.
% We summarize the costs for the four stable matchings of $\Pot_1$ and $\Pot_3$, two for each in table~\cref{tab:costs}.

%The formal definitions that we consider in this paper are described in \cref{sec:defi}

% \begin{table}[t!]
% \centering
% \begin{tabular}{lllll}
%   \toprule
%   & $M_1^1$ & $M_2^1$ & $M_1^3$ & $M_2^3$\\
%   \midrule
%   \egalcostn & $2$ & $3$ & $0$ & $2$\\
%   \regretcostn & $1$ & $2$ & $0$ & $2$\\
%   % \partnercostn & $1$ & $2$ & $0$ & $2$\\
%   \bottomrule
% \end{tabular}
% \caption{The optimality costs of the stable matching for $\Pot_1$ and $\Pot_3$ in \cref{fig:example}.}\label{tab:costs}
% \end{table}

% The \partnercostn{} is a natural generalization of the \emph{sex-equality} measure used for \SM~\cite{GusfieldIrving1989}. 
We call the corresponding optimization problems in the roommates setting \ESR and \RSR (see~\cref{sec:defi} for the formal definitions) and in the marriage setting \ESM and \RSM.

\subsubsection{\Egalcostn{} and \regretcostn} When the input preferences do not have ties (but could be incomplete), \ESM and \RSM are solvable in $O(n^4)$~time~\cite{IrLeGu1987}.

For preferences with ties, both \ESM and \RSM become NP-hard~\cite{MaIrIwMiMo2002}. Thus, already in the bipartite case, it becomes apparent that allowing ties in preference lists makes the task of computing an optimal egalitarian matching much more challenging.
\citet{MarxSchlotter2010} showed, among other results, that \ESM and \RSM are fixed-parameter tractable when parameterized by the parameter ``sum of the lengths of all ties''.
\citet{HaIrIwMaMiMoSc2003} showed inapproximability results for each of both problems.
\citet{IrvManOMa2009} showed that finding a perfect stable matching in a \SM{} instance with ties is NP-hard, even when the length of each preference list is bounded by three and the ties occur only on one side.
It is obvious that in the same setting, finding a perfect stable matching with minimum regret two (note that we defined the regret cost to be the maximum rank of the partner of any agent) remains NP-hard. 

For \ESR, \citet{Feder1992b} showed that the problem is NP-hard even if the preferences are complete and have no ties, and gave a 2-approximation algorithm for this case. 
\citet{CsIrMa2016} studied \ESR for preferences with bounded length~$\ell$ and without ties. They showed that the problem is polynomial-time solvable if $\ell=2$, and is NP-hard for $\ell \geq 3$.
They also showed that finding an arbitrary stable matching for a \SR instance with ties is NP-hard for $\ell=3$.
This immediately implies that \RSR with ties is NP-hard even for regret cost at most three.

\subsubsection{Other costs}

\newcommand{\secostn}{\textsf{equal-cost}}
\newcommand{\balcostn}{\textsf{balance-cost}}
\newcommand{\excostn}{\textsf{exchange-cost}}
For \textsc{Stable Marriage}, there are two more measures on socially optimal stable matchings: It may be desirable to find
\begin{itemize}[-]
  \item a \myemph{sex-equal} stable matching, which minimizes absolute value of the difference between the sums of the ranks of one side (\emph{i.e.}, the \myemph{sex-equal cost} $\rho$), 
  \item a \myemph{balanced} stable matching, which minimizes the maximum of the sums of the ranks of one side (\emph{i.e.}\ the \myemph{balance cost~$\balcost$}), and
  \item a \myemph{min.\ man-exchange} stable matching~$M$, which minimizes the number~$\excost$ of man-exchange pairs; a pair of two men~$u$ and $u'$ is called a \myemph{man-exchange pair in $M$}, if  they prefer the respective agent's partner to their own partner:
$M(u') \succ_u M(u)$ and $M(u) \succ_{u'} M(u')$.
\end{itemize}
Accordingly, we define the \myemph{$\secostn$}, the \myemph{$\balcostn$}, the \myemph{$\excostn$} of a matching~$M$ as
\begin{align*}
  \secostn(M) & \coloneqq \Big|\sum_{(u,w)\in M} \rank_u(w)- \sum_{(u,w)\in M}\rank_w(u)\Big|\text{, }\\
  \balcostn(M) & \coloneqq \max\Big(\sum_{(u,w)\in M} (\rank_m(w), \sum_{(u,w)\in M}\rank_w(u)\Big)\text{, and}\\
  \excostn(M) & \coloneqq |\{\{u,u'\} \subseteq U \mid M(u') \succ_u M(u) \wedge M(u) \succ_{u'} M(u')\}|.
\end{align*}
We call the corresponding optimization problems \SESM problem~\cite{Kato1993,MDIrv2014}, \BSM~\cite{Feder1992a}, and \MESM~\cite{Irving2008,DerCheSu2007}.
All three problems are NP-complete, even when ties are not present~\cite{Kato1993,Feder1992a,Irving2008}.
%The NP-hardness of \PSR follows by a similar proof for NP-hardness of \ESR; the later is given by \citet{Feder1992b}.

\citet{MDIrv2014} showed \SESM is NP-complete, even if no ties are present and the preference list of each agent has length at most three. 
When ties are present, its optimization variant cannot be approximated within arbitrary constant factor unless P${}={}$NP~\cite{HaIwMiYa2007}.

When ties are not present, with respect to the parameter~$t$, which is ``the balance cost~$\balcost$ plus the number of matched men'', it is straight-forward to verify that \BSM{} is fixed-parameter tractable.
Thus, \citet{GuRoSaZe2017} studied parameterized complexity of \BSM{} with respect to two parameters that measure that distance to the parameter~$t$.
They showed that for the first one it admits a kernel and for the second one it is W[1]-hard.
%\citet{CsIrMa2016} studied \SR and \ESR for preferences with bounded length~$\ell$ and without ties. They showed that both problems are polynomial-time solvable if $\ell=2$ and %each agent considers at most two other agents as acceptable.
%showed NP-hardness for $\ell=3$. That \SR is NP-hard for $\ell=3$ immediately implies that [\RSR is NP-hard even for regret cost at most three.

\citet{DerCheSu2007} showed that \MESM{} is NP-complete even without ties and when the preference list of each agent has length at most four.

\subsection{Distance/closeness to stable matchings}
For the case when no stable matchings exist, the agents may still be satisfied with a matching that is close to being stable. Such closeness could be measured, for example,
\begin{compactitem}[-]
\item by the cardinality~$\sss$ of a stable matching,
\item by the number~$\da$ of agents whose exclusion (by deleting these agents and their presence in all preference lists) allows perfect stability, 
\item by the number~$\bp$ of blocking pairs, 
\item by the number~$\ba$ of blocking agents (agents involved in blocking pairs), or
\item by the number~$\po$ of agents that prefer this matching to any other matching.
\end{compactitem}
The corresponding problems regarding the above measurements are denoted as \SRM, \SRD,  \SRB, and \SRA~(see \cref{sec:defi} for the formal definitions). We now list results regarding the complexity of these optimization problems.

\subsubsection{Cardinality} When the preferences may have ties, \SR becomes NP-hard~\cite{Ronn1990}, even if the input has complete preferences,
implying that \SRD, \SRM, and \SRB are all NP-hard in this case as well.

The closely related \SMM (\textsc{Max-SM}) problem (where the preferences are incomplete and have ties)
is NP-hard, even if 
\begin{inparaenum}[(1)]
  \item the ties are at the tails of the lists and occur on one side only,
  \item each list has at most one tie, and
  \item each tie is of length two~\cite{MaIrIwMiMo2002}.
\end{inparaenum}

\textsc{Max-SM}  has been extensively studied:
When on one side, the preference list of each agent has at most two agents, \citet{IrvManOMa2009} showed that \textsc{Max-SM}  can be solved in polynomial time by using a simple extension of the Gale/Shapley algorithm for the case without ties.
They compliment this tractability result by showing that even if each agent finds at most three agents acceptable, \textsc{Max-SM}  is NP-complete.

There are both constant-factor approximation algorithms and inapproximability results for the problem~\cite{HaIrIwMaMiMoSc2003,HaIwMiYa2007,IrMa2007,MaIrIwMiMo2002,IwMiYa2014,IwMiYa2008}.
In particular, \citet{HaIwMiYa2007} showed that both the maximization and the minimization variant of \textsc{Max-SM}  is NP-hard to approximate within some constant factor.
The result holds even if the preference lists have bounded length, and 
there is at most one tie per list, and the ties occur on one side only.

\citet{MarxSchlotter2010} studied the parameterized complexity of \SMM{} for parameters that are related to ties and showed that the problem is W[1]-hard with respect to the number of ties of the given instance.
Recently, \citet{AdGuRoSaZe2018} studied parameterized complexity of \SMM and \SRM.
In particular, they showed that \SMM parameterized by the cardinality~$\sss$ of a maximum stable matching and \SRM parameterized by the size of a maximum matching are fixed-parameter tractable.

\subsubsection{Deleting agents}
Note that by our definition of stability, a matching which leaves two agents unmatched can never be stable so that a stable matching resulting from precluding $\da$ agents in general cannot guarantee a stable matching with cardinality $(n-\da)/2$.

For preferences without ties, \citet{Tan1991} proposed a polynomial-time algorithm that computes a perfect stable matching in a \SR{} instance without ties, after deleting the fewest possible number of agents in an SR instance, showing that \SRD is polynomial-time solvable.
When ties are allowed, since it is already NP-hard to decide whether an SR instance admit an arbitrary stable matching, \SRD with ties NP-hard even if $\da=0$.
\citet{CHHY18wp} showed that \SRD with ties is fixed-parameter tractable with respect to the dual parameter~``number of agents remained in the stable matching''.

\subsubsection{Blocking pairs and blocking agents}\citet{AbBiMa2005} showed that \SRB is NP-hard, and cannot be approximated within a factor of $n^{0.5-\varepsilon}$ unless P${}={}$NP, even if the given preferences are complete. They also showed that the problem can be solved in $n^{O(\bp)}$~time, where $n$ and $\bp$ denote the number of agents and the number of blocking pairs, respectively.
This implies that the problem is in the parameterized complexity class XP for parameter~$\bp$. 
\citet{BiMaMcD2012} showed that the problem is NP-hard and APX-hard even if each agent has a preference list of length at most~$3$, and presented a $(2\ell-3)$-approximation algorithm for bounded list length~$\ell$. \citet{BiMaMi2010} and \citet{HaIwMi2009} showed that the related variant of \SM, where the goal is to find a matching with minimum blocking pairs among all maximum-cardinality matchings, cannot be approximated within $n^{1-\varepsilon}$ unless P${}={}$NP.

\citet{CHSYicalp-par-stable2018} showed that \SRB parameterized by the number~$\bp$ of blocking pairs and \SRA parameterized by the number~$\ba$ are W[1]-hard even if the length of the preference lists is bounded by five.

%\subsection{Other problem variants}

%The previous section deals with optimization variants of stable matching.
%In this section, we describe one notion on the quality of a matching and 

\subsubsection{Popular matchings}
\citet{Gaerdenfors1975} introduced the notion of \myemph{majority assignments} which is more widely known as \myemph{popular matchings}~\cite{AbrIrvKavMeh2007} and corresponds to the notion of weak Condorcet winners in voting theory~\cite{dCondorcet1785}.
Given two matchings,~$M_1$ and $M_2$, and an agent~$x$, we say that \myemph{$x$ prefers $M_1$ to $M_2$} if it holds that either~$x$ is matched in $M_1$ but unmatched in $M_2$ or $M_1(x) \succ_x M_2(x)$.
A matching~$M$ is called \myemph{popular} if for each matching~$M'$ it holds that the number of agents that prefer~$M$ to~$M'$ is \emph{no} less than the number of agents that prefer~$M'$ to~$M$.

For preferences without ties, one can verify that every stable matching is popular. 
Since a \SM{} instance always admits a stable matching,
every \SM{} is a yes instance for the popular matching question.
\citet{HuangKavitha2013} presented a polynomial-time algorithm which finds a popular matching with maximum cardinality in \SM{} for preferences without ties.
Very recently, \citet{GuMiSaZe2019soda,FaeKavPowZha2019soda} showed that in the roommate settings, even without ties, the problem of finding a popular matching is NP-hard.

For preferences with ties, \citet{AbrIrvKavMeh2007} showed that deciding whether a \SM{} instance admits a popular matching is NP-complete, but it is polynomial-time solvable when only one side has ties.
\citet{BirIrvMan2010} showed that for the roommates setting with ties,
both finding a perfect popular matching and finding a perfect popular matching are NP-hard.

%\clearpage
\section{Problem definitions}
\label{sec:defi}

\newenvironment{myquote}[1]%
  {\list{}{\leftmargin=0em\rightmargin=0.5em}\item[]}%
  {\endlist}

  \newcommand{\probDefSized}[4]{
  \begin{myquote}{#1}
   #2\\
  \textbf{Input:} #3\\[0.2ex]
  \textbf{Question:} #4
  \end{myquote}
}

\newcommand{\probDef}[3]{
  \probDefSized{2em}{#1}{#2}{#3}
}

\newcommand{\probOpt}[3]{
  \begin{myquote}{2em}
   #1\\
  \textbf{Input:} #2\\
  \textbf{Task:} #3
  \end{myquote}
}

\noindent Every problem described starting from \ref{min-bp-sr} is defined for the roommates setting without ties.
It can be restricted to the marriage case and generalized to the case with ties.
However, for some of the problems, we need to carefully adjust the measurement to also tackle the cost of agents that are unmatched. 

\begin{enumerate}
\item \probDef{\SM}
{Two disjoint sets~$U=\{u_1,u_2,\ldots, u_n\}$ and $W=\{w_1,w_2,\ldots, w_n\}$ of $n$ agents each, 
  and each agent~$u\in U$ (resp.\ $w\in W$) has a preference list~$\succ_{u}$ (resp.\ $\succ_{w}$) over $U$ (resp.\ over $W$).}
{Does $U\uplus W$ admit a stable matching?}

\item \probDef{\SR}
{A set~$U=\{u_1,u_2,\ldots, u_{2n}\}$ of $2n$ agents, 
  and each agent~$u\in U$ has a preference list~$\succ_{u}$ over (a subset of) $U$.}
{Does $U$ admit a stable matching?}

\item Variants: \SM with complete preferences but without ties (SM), \SM with complete preferences and with ties (SMT),  \SM with incomplete preferences and without ties (SMI), \SM with incomplete preferences and with ties (SMTI)

\item Variants: \SR with complete preferences but without ties (SR), \SR with complete preferences and with ties (SRT), \SR with incomplete preferences and without ties (SRI), \SR with incomplete preferences and with ties (SRTI)

\item\label{min-bp-sr} \probDef{\SRB~(\textsc{Min-BP-SR})}
{A \SR instance~$I$ and a number~$\bp\in \mathds{N}$.}
{Does $I$ admit a matching with at most $\bp$ blocking pairs?}

\item  \probDef{\SRA~(\textsc{Min-BA-SR})}
{A \SR instance~$I$ and a number~$\ba\in \mathds{N}$.}
{Does $I$ admit a matching with at most $\ba$ blocking agents?}

\item  \probDef{\SRM~(\textsc{Max-Size-SR})}
{A \SR instance~$I$ and a number~$\sss\in \mathds{N}$.}
{Does $I$ admit a stable matching with cardinality at least~$\sss$?}

\item  \probDef{\SRMin~(\textsc{Min-Size-SR})}
{A \SR instance~$I$ and a number~$\hat{\sss}\in \mathds{N}$.}
{Does $I$ admit a stable matching with cardinality at most~$\hat{\sss}$?}

\item  \probDef{\SRD~(\textsc{Del-SR})}
{A \SR instance~$I$ and a number~$\da\in \mathds{N}$.}
{Is there a matching which is stable for the instance obtained by deleting at most~$\da$~agents?}

\item  \probDef{\ESR (\textsc{Egal-SR})}
{A \SR instance~$I$ and a number~$\egalcost\in \mathds{N}$.}
{Does $I$ admit a stable matching~$M$ with \ecost{} at most $\egalcost$, \emph{i.e.}
  \[\sum_{u\in U} \rank_u(M(u)) \le \egalcost?\]}

Here, $\rank_u(M(u))$ is defined as the number of agents that are preferred to $M(u)$ by $u$.

\item  \probDef{\RSR (\textsc{Regret-SR})}
{A \SR instance~$I$ and a number~$\regretcost\in \mathds{N}$.}
{Does $I$ admit a stable matching with \rcost{} at most $\regretcost$,
  \emph{i.e.}\
  \[\max_{u\in U} \rank_u(M(u)) \le \regretcost?\]}

Here, $\rank_u(M(u))$ is defined as the number of agents that are preferred to $M(u)$ by $u$.

\item \probDef{\SESM (\textsc{SESM})}
{A \SM instance~$I$ with sets~$U$ and $W$, and a number~$\secost\in \mathds{N}$.}
{Does $I$ admit a stable matching with \secostn{} at most $\secost$,
  \emph{i.e.}\
  \[\Big|\sum_{(u,w)\in M} \rank_{u}(w) - \sum_{(u,w)\in M}\rank_w(u)\Big| \le \secost?\]}

\item \probDef{\BSM (\textsc{BSM})}
{A \SM instance~$I$ with sets~$U$ and $W$, and a number~$\balcost\in \mathds{N}$.}
{Does $I$ admit a stable matching with \balcostn{} at most $\balcost$,
  \emph{i.e.}\
  \[\sum_{(u,w)\in M} \rank_{u}(w) \le \balcost \text{ and } \sum_{(u,w)\in M}\rank_w(u) \le \balcost?\]}

\clearpage

\item \probDef{\MESM (\textsc{MESM})}
{A \SM instance~$I$ with sets~$U$ and $W$, and a number~$\excost\in \mathds{N}$.}
{Does $I$ admit a stable matching with \excostn{} at most $\excost$. % man-exchange pairs,
  \emph{i.e.}\
  \[ |\{\{u,u'\} \subseteq U\mid M(u') \succ_u M(u) \wedge M(u) \succ_{u'} M(u')\}| \le \excost?\]}
Here, a pair of two agents~$u,u'\in U$ is a \emph{man-exchange pair} if they prefer the respective agent's partner to their own partner:
$M(u') \succ_u M(u)$ and $M(u) \succ_{u'} M(u')$.

% \item \probDef{\DivSR~(\textsc{Divorce-SR})}
% {A \SR instance~$I$ with agent set~$U$, a matching~$M_0$ of $U$, and a number~$\doppar\in \mathds{N}$.}
% {Can matching~$M_0$ be transformed into a matching~$M_{\doppar}$ which is stable for $I$ via performing a sequence of $\doppar$~\dop{s}, 
%   \emph{i.e.}\ whether there is a sequence of pairs~$(p_0,p_1,\ldots, p_{\doppar})$  where for each pair~$p_i=\{u_i,w_i\}$, $0\le i \le \doppar-1$,
%  $M_{i+1}$ is recursively defined as $M_{i+1} \coloneqq M_{i} - \{u_i, M_i(u_i)\} - \{M_i(w_i), w_i\} + \{u_i, w_i\} + \{M_i(w_i), M_i(u_i)\}$ and
% the pair~$p_i$ is blocking matching~$M_{i}$, and
%  $M_\doppar$ is a stable matching for the given instance?}

\item \probDef{\textsc{Popular Matching in \SR}}
{A \SR instance~$I$ with agent set~$U$.}
{Does $I$ admit a popular matching~$M$, \emph{i.e.}
  \begin{align*}
    \forall \text{ matching } M'\colon |\{u\in U\mid M(u) \succ_u M'(u)\}| \ge |\{u\in U\mid M'(u) \succ_u M(u)\}|\text{?}
  \end{align*}
}
\end{enumerate}

%\bibliographystyle{abbrvnat}
%\bibliography{bib}

\end{document}